\documentstyle[12pt]{article}
\parskip=\medskipamount

\title{Green-Schwarz type  formulation of $D=11$ $S$ - invariant
superstring and superparticle actions.}
\author{A.A. Deriglazov\thanks{deriglaz@fma.if.usp.br ~ On leave of
    absence from Dept. of Math. Phys., Tomsk Polytechnical University,
    Tomsk, Russia}
~ and ~ D.M. Gitman\thanks{gitman@fma.if.usp.br}}
\date{Instituto de F\'\i sica, Universidade de S\~ao Paulo,\\
P.O. Box 66318, 05315-970, S\~ao Paulo, SP, Brasil.}

\begin{document}
\maketitle
\large
\begin{abstract}
A manifestly Poincare invariant formulations for $SO(1,10)$ and  
$SO(2,9)$ superstring
actions are proposed. The actions are invariant under a local fermionic
$\kappa$-symmetry as well as under a number of global symmetries, which
turn out to be on-shell realization of the known 
``new supersymmetry`` S-algebra. Canonical quantization of the theory
is performed and relation of the quantum state spectrum with that of
type IIA Green-Schwarz superstring is discussed. Besides, a mechanical
model is constructed , which is a zero tension limit of the $D=11$
superstring and which incorporates all essential features of the latter. 
A corresponding action 
invariant under off-shell closed realization of the S-algebra is obtained.
\end{abstract}
\noindent
{\bf PAC codes:} 0460D; 1130C; 1125

\section{Introduction}
 Green-Schwarz (GS) approach [1] to the construction of a manifestly 
Poincare invariant
actions for extended objects implies invariance under the local
$\kappa$-symmetry [2], which eliminates half of the initial fermionic
variables. It provides free dynamics for the physical sector variables
as well as supersymmetric spectrum of quantum states. The requirement
of consistency of the manifest super Poincare invariance with local
$\kappa$-symmetry leads to rather rigid restrictions on possible
dimensions of the target and worldvolume spaces in which the action can be
formulated.
These restrictions are enumerated in the known brane scan [3,4], which
prohibits, in particular, the Green-Schwarz type formulation for
$D=11$
 superstring action already at the
classical level. According to the brane scan the only permitted
dimensions are 3, 4, 6 and 10.

It would be an intriguing task to avoid this no-go theorem in relation
to recent progress in understanding of the eleven-dimensional nature of the
known superstring theories (see [4-9] and references therein).
In the strong coupling limit of  
M-theory $R^{11}\to \infty$, where $R^{11}$ is the radius of the 11th
dimension, the vacuum is eleven-dimensional Minkowski and the
effective field theory is $D11$ supergavity, which is viewed now as
strong coupling limit of ten-dimensional type IIA superstring [5].
Since $D11$ Poincare symmetry survives in this special point in
the moduli space of M-theory vacua, one may ask of the existence of 
a consistent $D11$ quantum theory in this limit (``uncompactified
M-theory'' according to Ref.[8]).
One possibility might be the supermembrane action [10-12], but in this
case one faces the problem of a continuous spectrum for the first
quantized supermembrane [13-15]. By analogy with the ten-dimensional case,
where the known field theories can be obtained as low energy limit of
the corresponding superstrings [16,8], a different natural candidate might
be a $D11$ superstring.

Several ways are known to avoid the no-go theorem, either by
considering space-time with non standard signature [17-20], or by
introducing higher spin worldvolume fields into the action
[22-28]\footnote{In recent works [29,30] $D=11$ action with second-class
constraints simulating a gauge fixation for the $\kappa$-symmetry was
 suggested. It was
achieved by adding of an appropriately chosen terms to the GS action
written in $D=11$. Since there is no $\kappa$-symmetry, identities of
the type (1), (2) are not necessary for the construction, but the
price is that only one half of supersymmetries survive in the
resulting Poincare invariant action. Supersymmetry of quantum states
spectrum for the model is under investigation now.}.
Since the brane scan is based on demanding of super Poincare
invariance, other possibility is to consider  $D=11$ GS type
superstring  actions for which the
supergroup is different from the super Poincare [4]. To elucidate how
it may work note that the crucial point of GS formulation 
for the case of superstring is the $\gamma$-matrix identity
\begin{equation}
\Gamma^\mu_{\alpha(\beta}(C\Gamma^\mu)_{\gamma\delta)}=0
\end{equation}
which holds in $D=3,4,6,10$.
It provides the existence of both global supersymmetry and local
$\kappa$-symmetry for the action [1].  
An eleven-dimensional analog of Eq.(1) has the form [3,31,32] 
\begin{equation}
 \Gamma^\mu_{\alpha(\beta}(C\Gamma^{\mu\nu})_{\gamma\delta)}+
(\Gamma^{\mu\nu})_{\alpha(\beta}(C\Gamma^\mu)_{\gamma\delta)} = 0,
\end{equation}
which contains antisymmetric product of
$\gamma$-matrices\footnote{Being appropriate for construction of
the supermembrane action [10],this identity does not allow one to
formulate $D=11$ super Poincare invariant action for superstring with
desirable properties. As was shown by Curtright [31], the globally
supersymmetric action based on this identity involves additional to
$x^i,\theta_a,\bar \theta_a$ degrees of freedom in the physical
sector. Moreover, it does not posses the $\kappa$-symmetry that could
provide free dynamics [31,33].}. It turns out to be applicable for the
superstring case instead of Eq.(1), if one replaces the standard
superspace 1-form
\begin{equation}
 dx^\mu - i\bar\theta \Gamma^\mu d\theta
\end{equation}
by an other one, which contains the same product of $\Gamma$-matrices as in
Eq.(2),
\begin{equation}
\ dx^\mu - i(\bar\theta\Gamma^{\mu\nu}d\theta)n_\nu.
\end{equation}
Appearance of the new variable $n^\mu (\tau,\sigma)$
seems to be an essential property of the construction
[17-20,29,30,34-36 ].
An action, which may be constructed from these 1-forms, is not
invariant  under the standard
supertranslations. As it will be shown the suitable generalization is
the ``new supersymmetry'' [18-20]
\begin{eqnarray}
\delta\theta =\epsilon, \qquad \delta
x^\mu=i(\bar\epsilon\Gamma^{\mu\nu}\theta)n_\nu.
\end{eqnarray}
The algebra of the corresponding generators is different from the super
Poincare and may be written as [17-20]
\begin{equation}
\{Q_\alpha,Q_\beta\}\sim \Gamma^{\mu\nu}P_\mu n_\nu.
\end{equation}
It is known as S-algebra previously discussed in the M-theory context
[17] (see [21] for discussion of the general case). 
To understand why it may be interesting, note that for the case of 
$SO(2,9)$ space with signature $(-,+,\cdots +,-)$ and in special 
Lorentz reference frame, where $n^\mu =(0,\cdots 0,1),$
eq.(5) reduces (see Appendix for our $\gamma$-matrix notations) to
the following one: 
\begin{equation}
\begin{array}{l}
\delta\theta^\alpha =\epsilon^\alpha, \qquad
\delta\bar\theta_\alpha =\bar\epsilon_\alpha,\\
\delta x^{\bar\mu}= -i\bar\epsilon_\alpha\tilde
\Gamma^{{\bar\mu}\alpha\beta}\bar\theta_\beta
+i\epsilon^\alpha\Gamma^{\bar\mu}_{\alpha\beta}\theta^\beta, \qquad
\delta x^{11}=0,
\end{array}
\end{equation}
where $\theta = (\bar\theta_\alpha,\theta^\alpha), \mu = (\bar\mu,11),
\bar\mu = 0,1,\cdots ,9.$ Equation (7) coincides with the standard
$D=10,N=2$ supersymmetry transformations. For the case of $SO(1,10)$
space with the standard signature, Eq.(5) reduces to $N=2$
supersymmetry up to a sign:$\left\{Q,Q\right\}\sim H, 
\left\{\bar Q,\bar Q\right\}\sim -H$. It may lead to a theory which is 
not manifestly unitary. The superalgebras with the ``wrong'' sign were
considered in recent work of Hull [40] where it was suggested that
the corresponding theories  are related with the standard ones by 
duality transformations. Both possibilities can be considered  
sumiltaneously, since our $D=11$ $\gamma$ - matrix notations are 
similar for these cases. Below, we discuss for definiteness the 
$SO(2,9)$ case. Thus, one can treat the new
supersymmetry (5) as a way to rewrite the $D=10, N=2$ supersymmetry in
 ``eleven dimensional notations'', and the corresponding  action
might be related to type IIA superstring. The possibility of lifting
the known ten-dimensional models to the manifestly invariant higher
dimensional form is under intensive investigation now 
[18-20,29,30,35,36], and the
main problem here is to find an appropriate Lagrangian formulation with
the variable $n^\mu$ treated on equal footing with all other ones. From
the previous discussion it is clear that the most preferable 
might be a formulation where the gauge $n^\mu=(0,\cdots ,0,1)$ would
be possible. Unfortunately, it is unknown how to introduce pure gauge
variable with the desired properties [18-20,34-36]. Below, we propose
superstring action, in which only zero modes of the auxiliary 
variables survive in the sector of physical degrees of freedom. 
Since the state spectrum of
a string is formed by the action on a vacuum of oscillator modes only,
one expects that the presence of zero modes for the case is not
essential. We demonstrate this fact within the canonical
quantization framework.

As compared to Refs.[18-20,35,36], an advantage of the present
formulation is that the explicit
Lagrangian action for $D=11$ superstring will be presented. Moreover,
since the variable $n^\mu (\tau,\sigma)$ is treated on equal footing
with other ones, global symmetry transformations form a superalgebra
in the usual sense, without appearance of nonlinear in generators terms
in the right hand side of Eq.(6) (see below). Thus, true form of the 
S-algebra will be obtained.

The work is organized as follows.
In Sec.2 the Hamiltonian analysis for the bosonic part of the $D=11$
superstring action is carried out. We show that the 
zero mode sector is
decoupled from $x$ sector. As a consequence, the
existence of the zero modes has no effect on the mass formula as well
as on the the spin content of the quantum states on each mass level,
which allows one to identify the corresponding part of the quantum 
state spectrum for the case of superstring with that of type
IIA superstring (let us point that the situation is similar
to the known relation between super $D$-string and type IIB 
superstring [37-39]).
In Sec.3 action of the $D=11$ superstring and its local
and global symmetries are presented. In Sec.4 we show that physical
degrees of freedom of the theory obey free equations of
motion. The canonical quantization of the theory and discussion of 
the state spectrum is presented. In Sec.5 zero-tension limit of the 
superstring action is
studied. We present $D=11$ action for mechanical system, which is invariant
under local $\kappa$-symmetry as well as under off-shell closed
realization of S-algebra of global symmetries. In the result, a
model-independent form of the S-algebra will be presented. Appendix
contains our $SO(2,9)$ $\gamma$-matrix conventions.

\section{Bosonic part of the action and its spectrum.}

As was mentioned in the Introduction, we need to get in our disposal
an auxiliary time-like vector variable. As a preliminary step to such
a construction we discuss $SO(2,D-2)$ action of the bosonic string 
modified by some additional terms with the above mentioned variable. 
The aim of this section is to show 
that the additional terms describe trivial degrees of freedom. 
An action for the $D=11$ superstring 
will be obtained in the next section as a supersymmetrization of the 
above mentioned bosonic action.

The action which will be examined is 
\begin{eqnarray}
S=\int d^2\sigma\left\{\frac{-g^{ab}}{2\sqrt{-g}}
\partial_a x^\mu\partial_b x^\mu -
\epsilon^{ab}\xi_a(n^\mu \partial_b x^\mu) - 
n^\mu\epsilon^{ab}\partial_a A^\mu_b -
\phi(n^2+1)\right\}.
\end{eqnarray}
Here $n^\mu(\sigma^a)$ is $D11$ vector and $d2$ scalar,
$A^\mu_a(\sigma^b)$ is $D11$ and $d2$ vector, while $\phi(\sigma^a)$
is a scalar. In Eq.(8) we have set $\epsilon^{ab} =
-\epsilon^{ba},\epsilon^{01}=-1$ and it also supposed that all the
variables are periodic on the interval $\sigma \subset [0,\pi]$
functions. From the equation of motion $\delta S/\delta\phi=0$ it
follows that $n^\mu$ is a time-like vector. 

Let us discuss the dynamics of the model. For this aim the Hamiltonian
formalism seems to be the most appropriate, since second-class
constraints will arise and must be taken into account.The total 
Hamiltonian constructed by means of standard procedure [41,42] has the form
\begin{eqnarray}  
H=\displaystyle\int d\sigma\left\{-\frac N2[\hat p^2+(\partial_1 x)^2]
-N_1(\hat p\partial_1 x)-\xi_0(n\partial_1 x)+(n\partial_1 A_0)+\right. \cr
\left.+\displaystyle\phi(n^2+1)+\omega^{ab}(\pi_g)_{ab}+
\lambda_\phi\pi_\phi+\lambda_{\xi a}{\pi_\xi}^a+\lambda^\mu_0 p^\mu_0+
\lambda^\mu_1(p^\mu_1-n^\mu)+\lambda^\mu_n p^\mu_n\right\},\label{ham}
\end{eqnarray}
where 
\begin{equation}
\hat p^\mu\equiv\ p^\mu+\xi_1 n^\mu, \qquad N\equiv\frac{\sqrt{-g}}{g^{00}},
 \qquad N_1\equiv\frac{g^{01}}{g^{00}},
\end{equation}

\noindent and $p^\mu$, $p^\mu_a$, $p^\mu_n$, $(\pi_g)_{ab}$, 
$\pi^a_\xi$, $\pi_\phi$
are momenta conjugated to the variables $x^\mu$, $A^\mu_a$, $n^\mu$,
 $g^{ab}$,
$\xi_a$, $\phi$ respectively; $\lambda_*$ are Lagrange multipliers
corresponding to the primary constraints.
The complete set of constrains can be found and presented as follows
\begin{equation}
 \qquad p^\mu_n=0, \qquad n^\mu-p^\mu_1=0;
\end{equation}
\begin{equation}
 \qquad \pi^1_\xi=0, \qquad \xi_1-(p_1 p)=0;
\end{equation}
\begin{equation}
 \qquad (\pi_g)_{ab}=0, \qquad \pi_\phi=0, \qquad \pi^0_\xi=0,
\qquad p^\mu_0=0;
\end{equation}
\begin{equation}
 \qquad (p_1)^2=-1, \qquad \partial_1 p^\mu_1=0;
\end{equation}
\begin{equation}
 \qquad H_0\equiv(p_1\partial_1 x)=0, \qquad H_\pm\equiv(\hat p^\mu\pm
\partial_1 x^\mu)^2=0;
\end{equation}
Constraints (11),(12)
are of second-class, while the remaining ones are first-class. An 
appropriate gauge fixing for the constraints (13) is
\begin{equation}
 \qquad g^{ab}=\eta^{ab}, \qquad \phi=\frac12, \qquad \xi_0=0,
 \qquad A^\mu_0=\displaystyle\int\limits_0^\sigma d\sigma'\xi_1\hat p^\mu.
\end{equation}
After introducing of Dirac brackets, which correspond to second-class set
(11)-(13),(16), the canonical pairs of variables $(n^\mu,p^\mu_n),
(\xi_a,\pi^a_\xi), (g_{ab}, {(\pi_g)}_{ab}), \\  
(\phi,\pi_\phi), (A^0_\mu,p^\mu_0)$ can be omitted. The Dirac brackets 
for the remaining variables coincide with the Poisson ones.
The choice in (16) simplifies the subsequent analysis of
$(A^\mu_1, p^\mu_1)$-sector, since the Hamiltonian equations of motion
for these variables look now as
\begin{equation}
 \qquad \partial_0 A^\mu_1=p^\mu_1, \qquad \partial_0 p^\mu_1=0.
\end{equation}
In order to find an appropriate gauge fixing for the constraints (14) let us
consider Fourier decomposition of periodical in the interval
$\sigma\subset[0,\pi]$ functions
\begin{equation}
\begin{array}{l}
 \qquad A^\mu_1(\tau,\sigma)=Y^\mu(\tau)+\sum\limits_{n\ne0} y^\mu_n(\tau)
e^{i2n\sigma},\\
 \qquad p^\mu_1(\tau,\sigma)=P^\mu_y(\tau)+\sum\limits_{n\ne0}
p^\mu_n(\tau) e^{i2n\sigma}.
\end{array}
\end{equation}
Then the constraint $\partial_1 p^\mu_1=0$ is equivalent to $p^\mu_n=0$,
$n\ne0$, and an appropriate gauge condition  is $y^\mu_n=0$, or,
equivalently, $\partial_1 A^\mu_1=0$. Thus, physical degrees of 
freedom in the sector $(A^\mu_1, p^\mu_1)$ are zero modes of these 
variables and the corresponding dynamics is
\begin{equation}
\begin{array}{l}
 \qquad A^\mu_1(\tau,\sigma)=Y^\mu+P^\mu_y\tau,\\
 \qquad p^\mu_1(\tau,\sigma)=P^\mu_y=const, \qquad (P_y)^2=-1.
\end{array}
\end{equation}
Since there are no of oscillator variables, this sector of the theory
may be considered as describing a point-like object, which propagates
freely according to Eq.(19). 
Dynamics of the remaining variables is governed now by the equations
\begin{equation}
 \qquad \partial_0 x^\mu=-p^\mu-(P_y p)P^\mu_y,
 \qquad\partial_0p^\mu=-\partial_1\partial_1 x^\mu.
\end{equation}
In addition, the constraints
\begin{equation}
 \qquad H_0\equiv(P_y\partial_1 x)=0,
 \qquad H_\pm\equiv(p^\mu+(P_y p)P^\mu_y\pm\partial_1 x^\mu)^2=0,
\end{equation}
hold, which obey the following algebra
\begin{equation}
\begin{array}{l}
 \qquad \left\{H_\pm,H_\pm\right\}=\pm4[H_\pm(\sigma)
\pm(P_y p)H_0(\sigma)+(\sigma\to\sigma')]
 \partial_\sigma \delta(\sigma-\sigma'),\\
 \qquad \left\{H_+,H_-\right\}=4[(P_y p)H_0(\sigma)+(\sigma\to\sigma')]
\partial_\sigma\delta(\sigma-\sigma'),\\
 \qquad \left\{H_0,H_\pm\right\}=\pm2 H_0(\sigma') \partial_\sigma
 \delta(\sigma-\sigma').
\end{array}
\end{equation}
On the $D=10$ hyperplane selected by the constraint $H_0(\sigma)=0$ it
reduces to the standard Virasoro algebra. Note also that the variable
$x^\mu(\tau,\sigma)$ obeys the free equation
$(\partial^2_\tau-\partial^2_\sigma) x^\mu=0$ 
as a consequence of Eqs.(20),(21).

To proceed further it is useful to impose the gauge condition 
\begin{equation}
(P_y\partial_1 p)=0,
\end{equation}
to the constraint $H_0=0$. By virtue of Eqs.(20),(23) one finds,
in particular, that $(P_y p)=(P_y P)$, where $P^\mu$ is the zero mode
of $p^\mu(\tau,\sigma)$. Then the solution to Eq.(20) (for the
case of closed world sheet) reads
\begin{equation}
\begin{array}{l}
 \qquad x^\mu(\tau,\sigma)=X^\mu-
\frac 1{\pi}(P^\mu+(P_y P)P^\mu_y)\tau+\\ \qquad \qquad \qquad
\frac i{2\sqrt\pi}\sum\frac 1n[\bar\alpha^\mu_n e^{i2n(\tau+\sigma)}
+\alpha^\mu_{-n} e^{-i2n(\tau-\sigma)}],\\
\qquad p^\mu(\tau,\sigma)=\frac 1{\pi} P^\mu+
\frac 1{\sqrt\pi}\sum[\bar\alpha^\mu_n
e^{i2n(\tau+\sigma)}-\alpha^\mu_{-n} e^{-i2n(\tau-\sigma)}],
\end{array}
\end{equation}
which is accompanied by the constraints
\begin{equation}
 \qquad P^\mu_y\bar\alpha^\mu_n=0, \qquad P^\mu_y\alpha^\mu_{-n}=0,
\end{equation}
\begin{equation}
\begin{array}{l}
 \qquad H_+=\frac{8}{\pi}\sum\limits^\infty_{-\infty}L_n
e^{i2n(\tau-\sigma)},
 \qquad L_n\equiv\frac12\sum\limits^\infty_{-\infty}
\alpha^\mu_{n-k} \alpha^\mu_k=0,\\
 \qquad H_-=\frac{8}{\pi}\sum\limits^\infty_{-\infty}\bar L_n
e^{i2n(\tau+\sigma)},
 \qquad \bar L_n\equiv\frac12\sum\limits^\infty_{-\infty}\bar
\alpha^\mu_{n-k}\bar\alpha^\mu_k=0,
\end{array}
\end{equation}
where $\alpha^\mu_0=-\bar\alpha^{\mu}_0\equiv\frac{1}{2\sqrt{\pi}}(P^\mu
+(P_y P)P^\mu_y)$.

>From Eq.(25) and the equality $(P^\mu+(P_y P)P^\mu_y)P^\mu_y=0$ for the
momenta of the center of mass, it follows that the sector $(x^\mu,p^\mu)$
of the theory describes, in fact, a closed string, which lives on
the (D-1)-dimensional hyperplane of standard signature which is 
orthogonal to the $P^\mu_y$ - direction.

Consider the following combinations:
\begin{equation}
\begin{array}{l}
 \qquad {\tilde X}^\mu\equiv x^\mu-\frac12\frac{(P_y Y)}{(P_y p)} P^\mu_y=
 {\bf X}^\mu-\frac1{\pi}{\bf P}^\mu_\tau +(oscillators),\\
 \qquad {\tilde P}^\mu\equiv p^\mu+(P_y p)P^\mu_y={\bf P}^\mu+
(oscillators),\\
 \qquad {\bf X}^\mu\equiv X^\mu-\frac12\frac{(P_y Y)}{(P_y P)}P^\mu_y,
 \qquad {\bf P}^\mu\equiv P^\mu+(P_y P)P^\mu_y,
\end{array}
\end{equation}
where solution of equations of motion (19), (24) was used. The 
quantities ${\bf X}^\mu, {\bf P}^\mu$ obey the Poisson brackets
$$
 \qquad \left\{{\bf X}^\mu,{\bf P}^\nu\right\}=\eta^{\mu\nu},
 \qquad \left\{{\bf X}^\mu,{\bf X}^\nu\right\}=\left\{{\bf P}^\mu,
{\bf P}^\nu\right\}=0.
\eqno{(27.a)} $$
\addtocounter{equation}{1}
and the same is true for ${\tilde X}^\mu,{\tilde P}^\mu$ quantities. As
 a consequence, the conserved charges:
$$
\begin{array}{l}
 \qquad {\bf P}^\mu=\displaystyle\frac1{\pi}\int_0^\pi d\sigma
 {\tilde P}^\mu, \\
 \qquad {\bf L}^{\mu\nu}=\displaystyle\frac1{\pi}\int_0^\pi d\sigma 
\tilde X^{[\mu}\tilde P^{\nu]}={\bf X}^{[\mu}{\bf P}^{\nu ]}
+{\bf S}^{\mu\nu}+\tilde{\bf S}^{\mu\nu},
\end{array}
\eqno{(28)} $$
where
$$
 \qquad {\bf S}^{\mu\nu}=i\sum_{n=1}^\infty(\alpha^\mu_{-n}\alpha^\nu_n-
\alpha^\nu_{-n}\alpha^\mu_n), \qquad 
\tilde{\bf S}^{\mu\nu}=i\sum_{n=1}^\infty(\bar\alpha^\mu_{-n}
\bar\alpha^\nu_n-\bar\alpha^\nu_{-n}\bar\alpha^\mu_n),
\eqno{(28.a)} $$
are generators of the Poincare group. 
This allows one to obtain the standard mass formulae for physical 
states. We adopt the Gupta-Bleuler prescription by requiring that 
physical states be annihilated by half of the operators 
$:L_n:$ , $:\bar L_n:$ 
\begin{equation}
(L_n-a\delta_{n,0})\mid phys >=(\bar L_n-a\delta_{n,0})
\mid phys >=0, \qquad n>0.
\end{equation}
By virtue of Eq.(26) for n=0 one finds the mass of the states
\begin{equation}
 \qquad m^2={\bf P}^2=-4\pi\left\{\sum\limits_{n>0}(\alpha^\mu_{-n}
\alpha^\mu_n+\bar \alpha^\mu_{-n}\bar \alpha^\mu_n)+2a\right\}.
\end{equation}
As it should be, the mass of a state is determined by oscillator 
excitations of $x^\mu(\tau,\sigma)$ --string only, zero modes of 
the sector $(A^\mu_1,p^\mu_1)$ do not make contribution
into this expression.

In order to describe the spectrum of the superstring suggested below,
it is useful to consider also noncovariant quantization in an
appropriately chosen coordinate system. By making use of a Lorentz
transformation one can consider coordinate system
where $P^\mu_y=(0,...0,1)$ (note that it is an admissible procedure in
the canonical quantization framework since the Lorentz transformation
is particular example of the canonical one). 
This breaks manifest $SO(2,D-2)$ covariance up to $SO(1,D-2)$ one.
In this basis Eq.(20)-(23) are reduced to
\begin{equation}
 \qquad \partial_0 x^{D-1}=0,  \qquad \partial_0 p^{D-1}=0;
\end{equation}
\begin{equation}
 \qquad \partial_0 x^{\bar\mu}=-p^{\bar\mu},
 \qquad \partial_0 p^{\bar\mu}=-\partial_1\partial_1 x^{\bar\mu},
 \qquad (p^{\bar\mu}\pm\partial_1 x^{\bar\mu})^2=0;
\end{equation}
where $\mu=(\bar\mu,D-1)$. Thus, zero modes of the theory (8) along the
direction $P^\mu_y$ decouples from $(D-1)$-dimensional sector (32),
while oscillator modes along the direction $P^\mu_y$ are absent
as a consequence of the equations $(P_y\partial_1x)=(P_y\partial_1p)=0$.

Thus, we have clear picture of the classical dynamics for the
model(8). The bosonic D - dimensional theory (8) can be considered as
 describing a composite object. The sector of the auxiliary variables 
$(A^\mu_1,p^\mu_1)$ corresponds to a point-like object. 
The only physical degrees of freedom of the sector 
are zero modes $Y^\mu$, $P^\mu_y$,
which describe propagation of a free particle, see Eq.(19). 
The sector of variables $(x^\mu,p^\mu)$ describes the closed string
(32),(30), which lives on (D-1) - dimensional hyperplane 
of the standard signature which is orthogonal
to $P^\mu_y$ - direction (the constraints (15), which relate the particle 
and the closed string mean that the latter one has no component
of center of mass momenta as well as of oscillator excitations
in the $P^\mu_y$ - direction, see Eqs.(24),(25)).

Next let us look at the spectrum of the quantum theory. The ground state of
the full theory 
$\mid p_0,0,p_{y0} >=\mid p_0 > \mid 0 > \mid p_{y 0} >$ is a direct
product of vacua, corresponding to the sectors 
$(X^\mu, P^\mu)$, $(\alpha^\mu_n, \bar\alpha^\mu_n)$, 
$(Y^\mu,P_y ^\mu),$
which obey $P_y^2\mid p_{y 0}>=-\mid p_{y 0}>$, 
$P^\mu\mid p_0>=p_0^\mu\mid p_0 >$, 
$\alpha_n^\mu\mid 0>=\bar\alpha_n^\mu\mid 0>=0$
for $n>0$. The excitation levels are then obtained by acting with 
\mbox{$n<0$} oscillators on the ground state and looks as 
$\left\{\Pi \alpha^\mu_n\cdots\bar\alpha^\nu_m\cdots\mid p_0, 0>
\right\}\times \\ 
\mid p_{y 0} >$. 
>From Eqs. (28), (25) one notes that the spin content on each mass level
(30) coincides with that of the $(D-1)$-dimensional closed string [16].
In the result, from the mass formulae (30) and Eqs. (28), (25), (32)
it follows that the quantum state spectrum of the theory (8) can be
identified with that of the $(D-1)$-dimensional closed bosonic
string. One notes that zero modes $Y^\mu, P^\mu_y$ manifest themselves
in additional degeneracy of the continuous part of the energy spectrum
only.

\section{Action of D=11 superstring and its symmetries}

As the $D=11$ superstring action we propose the following supersymmetric 
version of (8):
\begin{eqnarray}
S=\int d^2\sigma\left\{\frac{-g^{ab}}{2\sqrt{-g}}\Pi_a^\mu\Pi_{b\mu}-
i\varepsilon^{ab}\Big(\partial_a x^\mu-\frac i2\bar\theta\Gamma^{\mu\nu}
n_\nu\partial_a\theta\Big)(\bar\theta\Gamma_\mu\partial_b\theta)-\right.\cr
\left.-\varepsilon^{ab}\xi_a(n_\mu\Pi^\mu_b)-n_\mu\varepsilon^{ab}
\partial_aA^\mu_b-\phi(n^2+1)\right\},
\end{eqnarray}
where $\theta$ is a 32-component Majorana spinor of $SO(2,9)$, $\xi_a$
is a $d=2$ vector and $\Pi^\mu_a\equiv \partial_a x^\mu
-i\bar\theta\Gamma^{\mu\nu}n_\nu \partial_a\theta$. The role of the
last two terms was discussed in the previous section. The third term is
crucial for existence of local $\kappa$-symmetry and, at the same time,
it provides a split of oscillator part of the coordinate 
$x^{11}(\tau,\sigma)$ from the physical sector.

Let us describe global symmetry structure of the action (33). Bosonic
symmetries are the $D=11$ Poincar\'e transformations in the standard
realization, and additional transformations with antisymmetric parameter
$b^{\mu\nu}=-b^{\nu\mu}$,
\begin{equation}
\begin{array}{l}
\delta_bx^\mu={b^\mu}_\nu n^\nu,\\
\delta_bA^\mu_a=-{b^\mu}_\nu\displaystyle\left(\varepsilon_{ab}
\frac{g^{bc}}{\sqrt{-g}}{\Pi_c}^\nu-\xi_an^\nu+i(\bar\theta\Gamma^\nu
\partial_a\theta)\right).\end{array}
\end{equation}
The following fermionic supersymmetry transformations also take
place:
\begin{eqnarray}
&& \delta\theta=\epsilon, \qquad
 \delta x^\mu=i\bar\epsilon\Gamma^{\mu\nu}n_\nu\theta,\\
&& \delta A^\mu_a=i\varepsilon_{ab}\displaystyle\frac{g^{bc}}{\sqrt{-g}}
\Pi_{c\nu}(\bar\epsilon\Gamma^{\mu\nu}\theta)-\frac 56(\bar\epsilon
\Gamma^{\nu\mu}\theta)(\bar\theta\Gamma_\nu\partial_a\theta)+\cr
&&\qquad\qquad +\displaystyle\frac 16(\bar\epsilon\Gamma_\nu\theta)
(\bar\theta\Gamma^{\nu\mu}\partial_a\theta).\nonumber
\end{eqnarray}
One can prove that the complete algebra of symmetry transformations is
on-shell closed up to the
equation of motion $\partial_an^\mu=0$ and up to the trivial transformations
$\delta A^\mu_a=\partial_a\rho^\mu$ (see Eq.(38) below) with field-dependent
parameter $\rho^\mu$, as it usually happens in component formulations of
supersymmetric models without auxiliary fields. In Sec.5 an off-shell
closed version of these transformations will be obtained for the case
of $D=11$ superparticle. The only nontrivial
commutator is\footnote{To elucidate relation between Eqs.(36) and (6)
let us point a simple analogy: algebra of the Lorentz generators
$M^{\mu\nu}=\frac 12(x^\mu p^\nu-x^\nu p^\mu)$ can be written either as
$[M^{\mu\nu},M^{\rho\sigma}]=\eta^{\mu\rho}M^{\nu\sigma}+\dots$ or
$[M^{\mu\nu},M^{\rho\sigma}]=-\eta^{\mu\rho}p^\sigma x^\nu+\dots\,$.
The second case may be considered as corresponding to Eq.(6).}
\begin{equation}
[\delta_{\epsilon_1},\delta_{\epsilon_2}]=\delta_b, \qquad
b^{\mu\nu}=-2i(\bar\epsilon_1\Gamma^{\mu\nu}\epsilon_2).
\end{equation}
Let us note that one needs to use Fierz identities (which is the same
as for $SO(1,10)$ case)
\begin{equation}
(\Gamma^\mu)_{\alpha(\beta}(C\Gamma^{\mu\nu})_{\gamma\delta)}+
(\Gamma^{\mu\nu})_{\alpha(\beta}(C\Gamma^\mu)_{\gamma\delta)}=0
\end{equation}
to prove invariance of (33) under transformations (35) as well as to
check Eq. (36) for $A^\mu_a$ variable.
A relation of Eq.(35) to the $D=10,N=2$ supersymmetry has been
described in the Introduction.

Local bosonic symmetries for the action (33) are $d=2$
reparametrizations (with the standard transformation lows for all the
variables except the variable $\phi$, which transforms as a density,
$\phi'(\sigma')={\rm det}(\partial\sigma'/
\partial\sigma)\phi(\sigma)$~), Weyl symmetry, and the following 
transformations
with parameters $\rho^\mu(\sigma^a)$ and $\omega_a(\sigma^b)$, 
\begin{equation}
\begin{array}{l}
\delta A^\mu_a=\partial_a\rho^\mu+\omega_a n^\mu, \qquad
\delta\phi=-\frac12 \epsilon^{ab}\partial_a\omega_b.
\end{array}
\end{equation}
These symmetries are reducible since their combination with 
parameters of a special form ($\omega_a=\partial_a\omega,
 \rho^\mu=-\omega n^\mu$) is a trivial symmetry, 
$\delta_\omega A^\mu_a=-\omega\partial_a n^\mu, \delta_\omega\phi=0$
(note that $\partial_a n^\mu=0$ is one of the equations of
motion). Thus, Eq.(38) includes 12 essential parameters, which
correspond to the primary first-class constraints $p^\mu_0=0,
\pi_\phi=0$ (see below).

The action is also invariant under a pair of local fermionic
$\kappa$-symmetries. To describe them let us consider the following ansatz:
\begin{eqnarray}
&& \delta\theta=\pm\Pi_{d\mu}S^\pm\Gamma^\mu\kappa^{\mp d}, \qquad
 \delta x^\mu=-i\delta\bar\theta\Gamma^{\mu\nu}n_\nu\theta,\\
&& \delta g^{ab}=8i\sqrt{-g}P^{\pm ca}(\partial_c\bar\theta S^\mp
\kappa^{\mp b}),\nonumber
\end{eqnarray}
where 
\begin{equation}
S^\pm=\frac 12(1\pm n_\mu\Gamma^\mu), \quad
\kappa^{\mp d}\equiv P^{\mp dc}\kappa_c, \quad
P^{\mp dc}=\frac 12\left(\frac{g^{dc}}{\sqrt{-g}}\mp \varepsilon^{dc}
\right).
\end{equation}
Note that on-shell (where $n^2=-1$) the operators ${S^\pm}_\alpha{}^\beta$
form a pair of projectors in $\theta$-space. Let us remember
also that the $d=2$ projectors $P^\pm$ obey the following properties:
$P^{+ab}=P^{-ba}$, $P^{\mp ab}P^{\mp cd}=P^{\mp cb}P^{\mp ad}$.
After tedious calculations with the use of these properties and the
Fierz identities (37) a variation of the action (33) under the
transformations (39) can be presented in the form
\begin{equation}
\delta S=-\varepsilon^{ab}\partial_an_\nu G^\nu_b+
(n^2+1)H+\varepsilon^{ab}(n_\mu\Pi^\mu_b)F_a,
\end{equation}
where
\begin{eqnarray}
&&{G_b}^\nu\equiv -i\varepsilon_{bc}\displaystyle
\frac{g^{cd}}{\sqrt{-g}}(\delta\bar\theta\Gamma^{\mu\nu}\theta)
\Pi_{d\mu}+\frac 12 (\delta\bar\theta\Gamma^{\mu\nu}\theta)
(\bar\theta\Gamma_\mu\theta)-\cr
&&\qquad -\displaystyle\frac 12(\delta\bar\theta\Gamma_\mu\theta)
(\bar\theta\Gamma^{\mu\nu}\partial_b\theta)+i\xi_b
(\delta\bar\theta\Gamma^{\mu\nu}\theta)n_\nu,\cr
&& H\equiv +i\displaystyle\frac{g^{ab}}{\sqrt{-g}}
(\partial_a\bar\theta\Gamma^\mu\tilde\kappa^\mp)\Pi_{b\mu},\\
&& F_a\equiv i\Big[\varepsilon_{ac}\displaystyle\frac{g^{cd}}
{\sqrt{-g}}(\partial_d\bar\theta\Gamma^\mu\tilde\kappa^\mp)n_\mu+
(\partial_a\bar\theta\tilde\kappa^\mp)\mp\cr
&&\qquad \mp 2\varepsilon_{ab}P^{\pm\,cd}(\partial_c\bar\theta\Gamma^\mu
\kappa^{\mp\,b})\Pi_{d\mu}\Big],\qquad
\tilde\kappa^\mp\equiv\Pi_{a\mu}\Gamma^\mu\kappa^{\mp a}.\nonumber
\end{eqnarray}
All terms in Eq.(41) can evidently be canceled by the corresponding
variations of the auxiliary fields,
\begin{equation}
\delta A^\nu_b=G^\nu_b, \qquad 
\delta\phi=H,  \qquad
\delta\xi_a=F_a.
\end{equation}

In the result, the eleven dimensional superstring action (33) is invariant
under the transformations (39),(43). Let us stress that three last 
terms in the action (33) turn
out to be essential for achieving this local $\kappa$-symmetry.
Since in Eq.(39) there appeared the double projectors ($S^\pm$ and 
$\Pi_{a\mu}\Gamma^\mu$) acting on the $\theta$-space, the total number
of essential parameters is $8+8$. 

As a cheek-up of our calculations note that after the substitution 
$n^\mu=(0,\cdots0,1)$ the equations (39) are reduced to the ten-dimensional 
$\kappa$-symmetry transformations of the GS superstring action 
\begin{eqnarray}
&& \delta\theta^\alpha=+P^{-cd}\Pi^{\bar\mu}{}_d \tilde\Gamma^{\bar\mu\,
\alpha\beta}\bar\kappa_{c\beta}, \qquad
\delta\bar\theta_\alpha=P^{+cd}\Pi^{\bar\mu}{}_d \Gamma^{\bar\mu}_
{\alpha\beta}{\kappa_c}^\beta,\cr
&& \delta x^\mu=i\theta^\alpha\Gamma^{\bar\mu}_{\alpha\beta}\delta
\theta^\beta-i\bar\theta_\alpha\tilde\Gamma^{\bar\mu\,\alpha\beta}
\delta\bar\theta_\beta,\\
&& \delta g^{ab}=8i\sqrt{-g}\{-P^{-ca}(\partial_c\bar\theta\kappa^{+b})+
P^{+ca}(\partial_c\theta\bar\kappa^{-b})\}.\nonumber
\end{eqnarray}

\section{Dynamics of the D=11 superstring and D=10 type IIA GS
superstring.}

In this Section we are going to demonstrate that the dynamics of
physical variables in
the theory (33) is governed by the free equations. In the coordinate
system, where $n^\mu=(0,\cdots 0,1)$, the variables and the
corresponding equations can be identified with the ones of type IIA GS
superstring (modulo center of mass type variables $(Y^\mu, P^\mu_y)$
discussed in Sec.2). This conclusion is independent 
on the frame chosen since the initial action has $D=11$ Poincare 
invariance.

Performing the standard Hamiltonian analysis for the theory (33), one 
finds a pair of second-class constraints $p^\mu_n=0, ~
p^\mu_1-n^\mu=0$ among
primary constraints of the theory. Then the variables $(n^\mu,p^\mu_n)$ can
be omitted after introducing the associated Dirac bracket (see Sec. 2).
The Dirac brackets for the remaining variables coincide with the
Poisson ones, and the total Hamiltonian may be written as
\begin{eqnarray}
& H=\displaystyle\int d\sigma^1\left\{-\frac N2(\hat p^2+\Pi_{1\mu}
\Pi_1^\mu)-N_1\hat p_\mu \Pi_1^\mu +p_{1\mu}\partial_1 A_0^\mu
-\xi_0(p_{1\mu}\partial_1x^\mu)+\right. \cr
& \left.+\displaystyle\phi (p_1^2+1)+ \lambda_\phi\pi_\phi +
\lambda_{0\mu}p_0^\mu +\lambda^{ab}(\pi_g)_{ab}+\lambda_{\xi a} 
{p_\xi}^a+L_\alpha{\lambda_\theta}^\alpha\right\},\label{ham}
\end{eqnarray}
where $p^\mu, p^\mu_0, p^\mu_1, p_\xi^a, (\pi_g)_{ab}$ are momenta
conjugated  to the variables $x^\mu, A^\mu_0, A^\mu_1,\\ \xi_a,g_{ab}$
 respectively; $\lambda_*$ are Lagrange multipliers corresponding to
the primary constraints, and the following notations are used
\begin{equation}
\begin{array}{c}
N = \displaystyle\frac{\sqrt{-g}}{g^{00}}, \quad N_1=\frac{g^{01}}{g^{00}},
\quad \hat p^\mu=p^\mu-i\bar\theta\Gamma^\mu \partial_1\theta+
\xi_1p_1^\mu, \\
L_\alpha\equiv \bar p_{\theta\alpha}-i\displaystyle\Big(p_\mu-\frac i2
\bar\theta\Gamma_\mu\partial_1\theta\Big)\Big(\bar\theta\Gamma^{\mu\nu}\Big)
_\alpha p_{1\nu} \\
-i\Big(\partial_1x^\mu-\frac i2 \bar\theta\Gamma^{\mu\nu}p_{1\nu}
\partial_1\theta\Big)\Big(\bar\theta\Gamma_\mu\Big)_\alpha=0.\end{array}
\end{equation}
Poisson brackets for the fermionic constraints are: 
\begin{equation}
\left\{L_\alpha, L_\beta\right\}=2i\left [(\hat p^\mu+\Pi^\mu_1)
{(C\Gamma^\mu S^+)}_{\alpha\beta}-
(\hat p^\mu-\Pi^\mu_1){(C\Gamma^\mu S^-)}_{\alpha\beta}\right ],
\end{equation}
from which it follows that half of the latter are of first-class. The complete
system of constraints can be presented in the form 
$$
p_{\xi 1}=0, \qquad
\xi_1-(p p_1)+i(\bar\theta\Gamma^\mu\partial_1\theta)p_{1\mu}=0;
\eqno{(48.a)}$$
$$
{(\pi_g)}_{ab}=0, \qquad \pi_\phi=0, \qquad p_{\xi 0}=0, \qquad p^\mu_0=0;
\eqno{(48.b)}$$
$$
\partial_1 p^\mu_1=0, \qquad {(p^\mu_1)}^2=-1;
\eqno{(48.c)}$$
$$
H_0\equiv\partial_1 x^\mu p_{1 \mu}=0, \qquad 
H_\pm\equiv{(\hat p^\mu\pm\Pi^\mu_1)}^2=0, \qquad L_\alpha=0.
\eqno{(48.d)}$$
\addtocounter{equation}{1}
Besides, some equations for the Lagrange multipliers have been
determined 
in the course of Dirac procedure,
\begin{equation}
\lambda^\mu_n=0, \qquad
\lambda^\mu_1=\partial_1 A^\mu_0+2\phi p^\mu_1+Q^\mu;
\end{equation}
\begin{eqnarray}
(\hat p_\mu-\Pi_{1\mu})\Gamma^\mu S^-
(\lambda_\theta-\partial_1\theta)=0,\\
(\hat p_\mu+\Pi_{1\mu})\Gamma^\mu S^+
(\lambda_\theta+\partial_1\theta)=0;\nonumber
\end{eqnarray}
where
\begin{eqnarray}
& Q^\mu\equiv -N\xi_1\hat p^\mu-N_1\xi_1\Pi^\mu_1
-\xi_0\partial_1 x^\mu- \cr
& - \left [ ip_\nu\bar\theta\Gamma^{\nu\mu}+\frac12
(\bar\theta\Gamma_\nu\partial_1\theta)\bar\theta\Gamma^{\nu\mu}+
\frac12(\bar\theta\Gamma^{\nu\mu}\partial_1\theta)\bar\theta\Gamma_\nu\right
]\lambda_\theta;
\end{eqnarray}
and the Eq.(50) was obtained from the condition
$\left\{L_\alpha,H\right\}=0$. The constraints (48.a-c) were considered in
Sec.2
and we do not repeat the corresponding analysis here. Doing the gauge fixing
(16) and solving of the $(A^\mu_1,p^\mu_1)$-sector similar to the Eq.(19),
one can see that the dynamics
of the remaining variables is governed by equations of motion of the form
\begin{equation}
\begin{array}{l}
\qquad \partial_0 x^\mu=-(p^\mu+(pP_y)P^\mu_y)
-i(\bar\theta\Gamma^{\mu\nu}\lambda_\theta)P_{y\nu},\\
\qquad \partial_0 p^\mu=-\partial_1\left [ \partial_1 x^\mu
-i(\bar\theta\Gamma^{\mu\nu}\partial_1\theta)P_{y\nu}+
i\bar\theta\Gamma^\mu\lambda_\theta\right ],\\
\qquad \partial_0\theta^\alpha=-\lambda^\alpha_\theta,
\end{array}
\end{equation}  
together with the constraints (48.d). Equations for  
$\bar p_\theta$-variables are omitted since they are a consequence of
the constraints $L_\alpha=0$ and other equations.

Similarly to GS superstring, physical variables of the theory (33)
obey free equations of motion. To demonstrate this let us consider
the following decomposition for $\theta$-variable, $\theta=\theta^+
+ \theta^-$, where $\theta^\pm$ are spinors of opposite 
S-chirality\footnote{In the basis where $n^\mu=P^\mu_y=(0,\cdots
,0,1)$ the S-chiral spinors $\theta^\pm$ can be identified with $D=10$
Majorana-Weyl spinors of opposite chirality
$\theta^+=(\bar\theta_\alpha,0), \theta^-=(0,\theta^\alpha)$. Also, in
this basis $S^\pm$-projectors commutes with the light-cone 
$\Gamma^\pm$-matrices (See Appendix).}
\begin{equation}
\theta^\pm\equiv S^\pm\theta, \qquad
S^\mp\theta^\pm=0.
\end{equation}
By virtue of Eq.(50), the last equation from (52) can be rewritten as
\begin{equation}
(\hat p_\mu+\Pi_{1\mu})\Gamma^\mu(\partial_0-\partial_1)\theta^+=0, \qquad
(\hat p_\mu-\Pi_{1\mu})\Gamma^\mu(\partial_0+\partial_1)\theta^-=0.
\end{equation}
Further, the following conditions
\begin{equation}
\Gamma^+\theta^+=0, \qquad \Gamma^+\theta^-=0,
\end{equation}
turn out to be an appropriate gauge fixing for the first-class constraints,
which can be extracted from the equations $L_\alpha=0$. Then $\Gamma^+
\lambda^\pm_\theta$-projections vanish,
$\Gamma^+\lambda^\pm_\theta=0$, while for 
$\Gamma^-\lambda^\pm_\theta$-projections
one finds as a consequence of Eq.(50),\footnote{From equation 
$B_\mu\Gamma^\mu\Psi=0$ subject to condition $\Gamma^+\Psi=0$ it follows,
in particular, that $B^+\Gamma^-\Psi=0.$}
\begin{equation}
\Gamma^-\lambda^+_\theta=-\Gamma^-\partial_1\theta^+, \qquad
\Gamma^-\lambda^-_\theta=\Gamma^-\partial_1\theta^-.
\end{equation}
Besides, the following identities
\begin{equation}
\begin{array}{l}
\qquad \bar\theta\Gamma^+\lambda_\theta=\bar\theta\Gamma^i\lambda_\theta=0,\\ 
\qquad (\bar\theta\Gamma^{+\mu}\lambda_\theta)P_{y\mu}=
(\bar\theta\Gamma^{i\mu}\lambda_\theta)P_{y\mu}=0,\\
\qquad (\bar\theta\Gamma^{+\mu}\partial_1\theta)P_{y\mu}=
(\bar\theta\Gamma^{i\mu}\partial_1\theta)P_{y\mu}=0,
\end{array}
\end{equation}
hold in the gauge (55), where $i=1,2,\cdots ,8,11.$

Thus, we have, in fact, a situation which is similar to $D=10$ GS
superstring, and the further analysis coincides with the well known
case [1,16]. Besides the zero modes $(Y^\mu, P^\mu_y), (X^{10},
P^{10})$  which was considered in Sec.2 and are fully decoupled from
the others,
physical variable sector contains the transverse components
$x^i, i=1,\cdots ,8$ of the coordinate $x^\mu$, and a pair
of 32-component spinors $\theta^\pm$ constrained by the equations (53),(55). 
By virtue of Eqs.(52)-(57) one gets that the physical variables obey
the free equations
\begin{equation}
\begin{array}{l}
\partial_0 x^i=-(p^i+(pP_y)P^i_y), \qquad 
\partial_0 p^i=-\partial_1\partial_1 x^i; \\
(\partial_0-\partial_1)\Gamma^-\theta^+=0, \qquad
(\partial_0+\partial_1)\Gamma^-\theta^-=0.
\end{array}
\end{equation}

To analyze the  quantum state spectrum for the theory under
consideration let us follow on
the SO(8) covariant procedure described in Sec.2. In the basis where
$P^\mu_y=(0,\cdots ,0,1)$ the gauge conditions (55) are
equivalent to $\Gamma^+\theta=0$ with the solution being 
$\theta=(\theta_a, 0, 0, \bar\theta_{\dot a})$, where $\theta_a,
\bar\theta_{\dot a}$ are SO(8) spinors of opposite chirality. Then the
second line in (59) reduces to $(\partial_0-\partial_1)\theta_a=0 ,
 (\partial_0+\partial_1)\bar\theta_{\dot a}=0$, while the second  
equation from (48.d) coincides  with the ten-dimensional Virasoro
constraints. 
General state of the theory is 
\begin{eqnarray}
\qquad [\Pi\alpha^\mu_n\cdots\bar\alpha^\nu_m\cdots S^a_k\cdots\bar S^b_p
\cdots |0, p_0 >]\otimes |p_{y0}>
\end{eqnarray}
where the bosonic $(\alpha)$ and fermionic $(S)$ oscillators are
identical to type IIA superstring oscillators. From analysis of the
mass formula and of the spin content on each mass level (which is 
similar to that of Sec.2) 
it follows, that the expression in square brackets of Eq.(59) can be
identified with the state spectrum of type IIA GS superstring.
One notes that zero modes 
$Y^\mu, P_y^\mu$ manifest themselves in additional degeneracy of the
continuous part of the energy spectrum only.
In conclusion, let us point analogy: recently [38] it was established
that super D-string is canonically equivalent to type IIB GS
superstring with $\Theta$-term added (the $\Theta$-term contains the
world-volume vector of D-string, and zero modes of the vector survive 
in the physical variable sector).
Equivalence in the path integral framework was established in
[39]. Situation with the theory under consideration is similar, but we
have $D=11$ action  related with type IIA superstring action.

\section{D=11 mechanical system with off-shell closed new supersymmetry
S-algebra.}

Being zero-tension limit of the GS superstring, the
Casalbuoni-Brink-Schwarz superparticle incorporates all its essential
features [44,45]. It allows one to study the problem in a
more simple framework of the mechanical model. In a similar fashion, in
this Section a point-like analog for the $D=11$ superstring is
presented and discussed. The action is invariant under local 
$\kappa$-symmetry as well as under a number of global symmetries with
on-shell closed algebra of commutators. Its off-shell closed version
will be obtained by a slight modification of the initial action, which
allows one to extract a true form of the S-algebra. Being
model-independent, it may be used now as a basis for systematic 
construction of various $D=11$ models.

Our starting point is the following $SO(2,9)$ Lagrangian action
\begin{equation}
\begin{array}{l}
S=\displaystyle\int d\tau\left\{\frac{1}{2e}\Pi^\mu\Pi_\mu+n^\mu\dot z^\mu-
\phi(n^2+1)\right\}, \\
\qquad \Pi^\mu\equiv\dot x^\mu-i(\bar\theta\Gamma^{\mu\nu}\dot\theta)n_\nu-
\xi n^\mu,
\end{array}
\end{equation}
with all the variables being functions on the evolution parameter $\tau$.
Note that the last two terms are, in fact, an action for bosonic particle 
$z^\mu(\tau)$ written in the first-order form.

Global bosonic symmetries of the action (60) are $D=11$ Poincare
transformations (with the variable $n^\mu$ being inert under the
Poincare shifts), and the following transformations 
\begin{equation}
\qquad\delta_b x^\mu=b^\mu{}_\nu n^\nu, \qquad 
\delta_b z^\mu=-\frac1{e}b^\mu{}_\nu\Pi^\nu,
\end{equation}
with antisymmetric parameter $\omega^{\mu\nu}=-\omega^{\nu\mu}$. There
is also a global symmetry with a fermionic parameter $\epsilon^\alpha$,
\begin{equation}
\delta_\epsilon\theta=\epsilon, \qquad
\delta_\epsilon x^\mu=-i(\bar\theta\Gamma^{\mu\nu}\epsilon)n_\nu, \qquad
\delta_\epsilon z^\mu=-\frac{i}{e}(\bar\epsilon\Gamma^{\mu\nu}\theta)\Pi_\nu.
\end{equation}
The algebra of the corresponding commutators turns out to be on-shell
closed and looks as follows:
\begin{eqnarray}
&& [\delta_{b1}, \delta_{b2}] x^\mu=0, \qquad
[\delta_{b1}, \delta_{b2}] z^\mu=\frac 1{e}{}b_1{}^\mu{}_\nu
(\delta_{b2}\Pi^\nu)-(1\leftrightarrow2);\cr
&& [\delta_{\epsilon 1}, \delta_{\epsilon 2}]\theta=0, \qquad
[\delta_{\epsilon 1}, \delta_{\epsilon 2}] x^\mu=\delta_b x^\mu,\\
&& [\delta_{\epsilon 1}, \delta_{\epsilon 2}] z^\mu=\delta_b z^\mu+
\left [\frac{i}{e} (\bar\epsilon_1\Gamma^{\mu\nu}\theta)
(\delta_{\epsilon 2}\Pi_\nu)-(1\leftrightarrow 2)\right ],
b^{\mu\nu}\equiv -2i(\bar\epsilon_1\Gamma^{\mu\nu}\epsilon_2);\cr
&& [\delta_\epsilon, \delta_b]\theta=0, \qquad
[\delta_\epsilon, \delta_b] x^\mu=0,\cr
&& [\delta_\epsilon, \delta_b] z^\mu=
-\frac 1{e} b^\mu{}_\nu(\delta_\epsilon\Pi^\nu)+
\frac{i}{e}(\bar\epsilon\Gamma^{\mu\nu}\theta)\delta_b\Pi_\nu.\nonumber
\end{eqnarray}
Commutators with the Poincare transformations are omitted here since they
have the standard form. All the extra terms in the right hand side of
Eq.(63) contain $\delta\Pi^\mu\sim\dot n^\mu$ and vanish on-shell,
where $\dot n^\mu=0$. To find off-shell closed version of these
transformations let us note that all extra terms arise owing to the
variation of the $\Pi^\mu$-term. The latter appears, in its turn, due
to variation of the variable $z^\mu$.
Following the standard ideology [46,47], these terms can be canceled by
replacing $\Pi^\mu\to(\Pi^\mu-B^\mu)$ in Eqs.(61),(62), where the
auxiliary variable $B^\mu$ has the same transformation properties as 
$\Pi^\mu$, $\delta B^\mu=\delta\Pi^\mu$. The resulting off-shell
closed
version of the global symmetries is 
\begin{eqnarray}
&&\delta_\epsilon\theta=\epsilon, \qquad
\delta_\epsilon x^\mu=-i(\bar\theta\Gamma^{\mu\nu}\epsilon)n_\nu,\cr
&&\delta_\epsilon z^\mu=
-i(\bar\epsilon\Gamma^{\mu\nu}\theta)
\left [ \frac 1{e}\Pi_\nu-B_\nu\right ], \qquad
\delta_\epsilon B^\mu=
\frac {i}{e} (\bar\epsilon\Gamma^{\mu\nu}\theta) \dot n_\nu;
\end{eqnarray}
\begin{equation}
\delta_b x^\mu=b^\mu{}_\nu n^\nu, \qquad
\delta_b z^\mu=-\omega^\mu{}_\nu( \frac 1{e}\Pi^\nu-B^\nu), \qquad
\delta_b B^\mu=\frac 1{e} \omega^\mu{}_\nu \dot n^\nu,
\end{equation}
while the final form of the action, which is invariant under these 
transformations, looks as follows:
\begin{equation}
\begin{array}{l}
S=\displaystyle\int d\tau\left\{\frac 1{2e} \Pi^\mu\Pi_\mu+
n^\mu \dot z^\mu-\phi(n^2+1)-\frac 12 B^2\right\}.
\end{array}
\end{equation}

Thus, S-algebra consist of Poincare subalgebra $(M^{\mu\nu}, P^\mu)$,
and includes generators of the new supertranslations $Q_\alpha$ as
well as second-rank Lorentz tensor $Z_{\mu\nu}$, corresponding to 
transformation (65). The only nontrivial commutator is
\begin{equation}
\left\{Q_\alpha, Q_\beta\right\}
=2i{(C\Gamma^{\mu\nu})}_{\alpha\beta}Z_{\mu\nu}.
\end{equation}
Note, that it is not a modification of the super Poincare algebra, but
essentially different one, since the commutator of the 
supertranslations leads  to $Z$-transformation instead of the Poincare
shift.

The action (66) is also invariant under the local $\kappa$-symmetry
transformations
\begin{eqnarray}
&& \delta\theta=\Pi_\mu\Gamma^\mu\kappa,\cr
&&\delta x^\mu=i(\bar\theta\Gamma^{\mu\nu}\delta\theta) n_\nu,\qquad
\delta z^\mu=
-\frac {i}{e}(\bar\theta\Gamma^{\mu\nu}\delta\theta)\Pi_\nu,\cr
&&\delta e=4ie(\dot{\bar\theta}\Gamma^\mu\kappa)n_\mu, \qquad
\delta\xi=-2i(\dot{\bar\theta}\delta\theta).
\end{eqnarray}
This fact is essential to confirm that physical sector variables
obey free equations of motion. The Hamiltonian analysis for the model 
is similar to that of the superstring action discussed above, and 
is as follows. One finds the total Hamiltonian
\begin{equation}
\begin{array}{l}
H=\displaystyle\frac {e}2 p^2+\xi(p p_z)+\phi(p^2_z+1)+\lambda_e\pi_e
+\lambda_\xi p_\xi+\lambda_\phi\pi_\phi+\lambda^\mu_B p_{B\mu}+\\
\qquad\lambda_{n\mu}p^\mu_n + \lambda_{z\mu}(p^\mu_z-n^\nu)
+ L_\alpha\lambda^\alpha_\theta,
\end{array}  
\end{equation}  
and the constraints
$$
p^\mu_n=0, \qquad p^\mu_z-n^\mu=0;
\eqno{(70.a)}$$
$$
\pi_e=0, \qquad \pi_\phi=0, \qquad p_\xi=0, \qquad  p^\mu_B=0;
\eqno{(70.b)}$$
$$
p^2_z=-1, \qquad (p p_z)=0, \qquad  p^2=0;
\eqno{(70.c)}$$
$$
L_\alpha\equiv\bar p_{\theta\alpha}-
i{(\bar\theta'\Gamma^\mu)}_\alpha p_\mu=0,
\eqno{(70.d)}$$
\addtocounter{equation}{1} 
where $\theta' \equiv p_{z\mu}\Gamma^\mu \theta$. The matrix
of the Poisson brackets of fermionic constraints
\begin{equation}
\left\{L_\alpha, L_\beta\right\}
=2i{(C\Gamma^{\mu\nu})}_{\alpha\beta} p_\mu p_{z\nu},
\end{equation}
is degenerated on the constraints surface as a consequence of the
identity 
${(\Gamma^{\mu\nu}p_\mu p_{z\nu})}^2=4[(p p_z)-p^2 p^2_z]{\bf 1}=0$.
It means that half of the constraints are first-class. Also, from 
the condition $\left\{L_\alpha, H\right\}=0$ one finds equation, which 
determine $\lambda_\theta$-multipliers, 
\begin{equation}
p_\mu\Gamma^\mu\lambda'_\theta =0, \qquad 
\lambda'_\theta \equiv p_{z\mu}\Gamma^\mu\lambda_\theta.
\end{equation}
After a gauge fixation for the first-class constraints (70.b) (and
take into account the second-class constraints (70.a)), the canonical
pairs $(e, \pi_e),\\
(\phi, \pi_\phi),(\xi, p_\xi),(B^\mu, p^\mu_B),
(n^\mu, p^\mu_n)$ can be omitted from the consideration. The dynamics 
of the remaining variables is governed by the equations
$$
\dot z^\mu=p^\mu_z+i(\bar\theta\Gamma^{\mu\nu}\lambda_\theta)p_\nu, \qquad
\dot p^\mu_z=0;
\eqno{(73.a)}$$
$$
\dot x^\mu=p^\mu-i(\bar\theta\Gamma^{\mu\nu}\lambda_\theta)p_{z\nu}, \qquad
\dot p^\mu=0;
\eqno{(73.b)}$$
$$
\dot \theta^\alpha=-\lambda^\alpha_\theta, \qquad
\dot{\bar p}_{\theta\alpha}=0.
\eqno{(73.c)}$$
\addtocounter{equation}{1}
The next step is to impose a gauge for the first-class constraints
which are contained among the equations (70.d),
\begin{equation}
\Gamma^+\theta'=0.
\end{equation}  
By virtue of (72),(73.c) all the $\lambda_\theta$-multipliers can
be determined, $\lambda_\theta=0$, and Eqs.(73.a-c) are reduced to
free equations.

The resulting picture corresponds to zero-tension limit of the $D=11$
superstring action (33). The above consideration of the physical
sector allows one to treat the system as a composite one.
It consist of the
bosonic $z^\mu$-particle (73.a) and the superparticle (73.b), (73.c), 
subjected to the constraints (70.c). Their free propagation is restricted
by the kinematic constraint $(p p_z)=0$, which means that the
superparticle lives on $D=10$ hyperplane of the standard signature which
is orthogonal to the direction of motion of $z^\mu$-particle.

\section{Conclusion.}

One can consider $D=10$ GS superstring action as a lift of 
SO(8)-covariant superstring formulation up to the manifestly 
SO(1,9)-invariant form. In this paper we have considered, in fact, 
the next step of such a  lift, from SO(1,9) up to SO(2,9) or
SO(1,10). The key 
point is that the action constructed is based
on the superalgebra of global symmetries (34)-(36), (67), which is
nonstandard super extension of the Poincare one. It allows one 
to avoid restrictions of the brane scan followed from demanding of 
the super Poincare invariance. In the result, we have constructed 
$N=1$ S-invariant action for $D=11$ superstring with the quantum 
state spectrum which can be identified with that of $D=10$, 
type IIA GS superstring. The
only difference is an additional infinite degeneracy in the 
continuous part of the energy spectrum, related with the zero modes
$Y^\mu, P_y^\mu$. On the classical level these degrees of freedom
may be identified with coordinate and momenta of a free propagating
point-like object.

In accordance with the results of Refs.[18] and [20] one expects that
critical dimension of the theory is $D=11$. We hope that similar
construction will work for lifting of the $D=10$ type IIB string to
the corresponding (10,2) version. It will be also 
interesting to apply the scheme developed in this work for construction 
of the Lagrangian formulation for $(D-2,2)$ SYM equations of motion
considered in [35,36].

\section*{Acknowledgments.}

One of the authors (A.A.D.) thanks N. Berkovits and J. Gates for
useful discussions.

D.M.G. thanks Brasilian foundation CNPq for permanent support. The
work of A.A.D. has been supported by the Joint DFG-RFBR project No
96-02-00180G, by Project INTAS-96-0308, and by FAPESP.

\section*{Appendix}
\setcounter{equation}{0}
\def\theequation{A.\arabic{equation}}
In this Appendix we describe the minimal spinor representation of the
Lorentz group $SO(2,9)$, which is known to have dimension $2^{[D/2]}$.
To this aim, it is enough to find eleven $32\times32$ matrices 
$\Gamma^\mu$  satisfying the anticommutation relations 
$\Gamma^\mu\Gamma^\nu+\Gamma^\nu\Gamma^\mu=
-2\eta^{\mu\nu}$, $\mu,\nu=0,1,\dots,9,11$, $\eta^{\mu\nu}=(-,+,\dots,+,-)$.
A convenient way is to use the well known $16\times 16$
$\gamma$-matrices of $SO(1,9)$ group, which we denote as
$\Gamma^m_{\alpha\beta}$, $\tilde\Gamma^{m\alpha\beta}$, $m=0,1,\dots,9$.
Their explicit form is:
\begin{eqnarray}
&& \Gamma^0=\left(\begin{array}{cc} {\bf 1}_8 & 0\\
0 & {\bf 1}_8\end{array}\right),\;\; 
 \tilde\Gamma^0=\left(\begin{array}{cc} -{\bf 1}_8 & 0\\
0 & -{\bf 1}_8\end{array}\right),\cr
&& \Gamma^i=\left(\begin{array}{cc} 0 & {\gamma^i}_{a\dot a}\\
\bar\gamma^i{}_{\dot aa} & 0 \end{array}\right), \;\;
 \tilde\Gamma^i=\left(\begin{array}{cc} 0 & {\gamma^i}_{a\dot a}\\
\tilde\gamma^i{}_{\dot aa} & 0\end{array}\right),\cr
&& \Gamma^9=\left(\begin{array}{cc} {\bf 1}_8 & 0\\
0 & -{\bf 1}_8\end{array}\right), \;\;
 \tilde\Gamma^9=\left(\begin{array}{cc} {\bf 1}_8 & 0\\
0 & -{\bf 1}_8\end{array}\right),
\end{eqnarray}
where ${\gamma^i}_{a\dot a}$, $\bar\gamma^i{}_{\dot aa}\equiv
({\gamma^i}_{a\dot a})^{\rm T}$ are real $SO(8)$ $\gamma$-matrices
\cite{16},
\begin{equation}
\gamma^i\bar\gamma^j+\gamma^j\bar\gamma^i=2\delta^{ij}{\bf 1}_8,
\end{equation}
and $i,a,\dot a=1,\dots,8$. As a consequence, the matrices $\Gamma^m$,
$\tilde\Gamma^m$ are real, symmetric, and obey the anticommutation relation
\begin{equation}
\{\Gamma^m, \tilde\Gamma^n\}=2\eta^{mn}{\bf 1},
\end{equation}
where $\eta^{mn}=(-,+,\dots,+)$. Then a possible realization for the $D=11$
$\gamma$-matrices is
\begin{equation}
\Gamma^\mu=\left\{\left(\begin{array}{cc} 0 & \Gamma^m\\
-\tilde\Gamma^m & 0\end{array}\right), \left(\begin{array}{cc}
{\bf 1}_{16} & 0\\ 0 & -{\bf 1}_{16}\end{array}\right)\right\},
\end{equation}
The properties of $\Gamma^m$,
$\tilde\Gamma^m$ induce the following relations for $\Gamma^\mu$:
\begin{eqnarray}
&& (\Gamma^0)^{\rm T}=\Gamma^0, \qquad
(\Gamma^i)^{\rm T}=-\Gamma^i, \qquad
{(\Gamma^{11})}^{\rm T}=\Gamma^{11} \cr
&&(\Gamma^\mu)^*=\Gamma^\mu, \qquad 
\{\Gamma^\mu,\Gamma^\nu\}=-2\eta^{\mu\nu}{\bf 1}_{32},
\end{eqnarray}
The charge conjugation matrix $C$,
\begin{eqnarray}
C\equiv\Gamma^0\Gamma^{11}, \qquad
C^{-1}=-C, \qquad
C^2=-{\bf 1}
\end{eqnarray}
can be used to construct the symmetric matrices $C\Gamma^\mu$, 
$(C\Gamma^\mu)^{\rm T}=C\Gamma^\mu$.
One can introduce antisymmetrized products
\begin{equation}
\Gamma^{\mu\nu}=\frac 12(\Gamma^\mu\Gamma^\nu-\Gamma^\nu\Gamma^\mu),
\end{equation}
which have the following explicit form in terms of the corresponding
$SO(1,9)$ and $SO(8)$ matrices:
\begin{eqnarray}
&& \Gamma^{0i}=-\left(\begin{array}{cc}
\Gamma^{0i} & 0\\ 0 & \tilde\Gamma^{0i}\end{array}\right)=\left(
\mbox{\begin{tabular}{c|c}
$\begin{array}{cc} 0 & -\gamma^i\\ -\bar\gamma^i & 0\end{array}$ & 0\\
\hline
0 & $\begin{array}{cc} 0 & \gamma^i\\ \bar\gamma^i & 0\end{array}$
\end{tabular}}\right),\cr
&& \Gamma^{09}=-\left(\begin{array}{cc}
\Gamma^{09} & 0\\ 0 & \tilde\Gamma^{09}\end{array}\right)=\left(
\mbox{\begin{tabular}{c|c}
$\begin{array}{cc} -1 & 0\\ 0 & 1\end{array}$ & 0\\
\hline
0 & $\begin{array}{cc} 1 & 0\\ 0 & -1\end{array}$
\end{tabular}}\right),\cr
&& \Gamma^{ij}=-\left(\begin{array}{cc}
\Gamma^{ij} & 0\\ 0 & \tilde\Gamma^{ij}\end{array}\right)=\left(
\mbox{\begin{tabular}{c|c}
$\begin{array}{cc} -\gamma^{ij} & 0\\ 0 & -\bar\gamma^{ij}\end{array}$ & 0\\
\hline
0 & $\begin{array}{cc} -\gamma^{ij} & 0\\ 0 & -\bar\gamma^{ij}\end{array}$
\end{tabular}}\right),\cr
&& \Gamma^{i9}=-\left(\begin{array}{cc}
\Gamma^{i9} & 0\\ 0 & \tilde\Gamma^{i9}\end{array}\right)=\left(
\mbox{\begin{tabular}{c|c}
$\begin{array}{cc} 0 & \gamma^i\\ -\bar\gamma^i & 0\end{array}$ & 0\\
\hline
0 & $\begin{array}{cc} 0 & \gamma^i\\ -\bar\gamma^i & 0\end{array}$
\end{tabular}}\right),\\
&& \Gamma^{0,11}=\left(\begin{array}{cc}
0 & -\Gamma^0\\ -\tilde\Gamma^0 & 0\end{array}\right)=\left(
\mbox{\begin{tabular}{c|c}
0 & $\begin{array}{cc} -{\bf 1} & 0\\ 0 & -{\bf 1}\end{array}$\\
\hline
$\begin{array}{cc} {\bf 1} & 0\\ 0 & {\bf 1}\end{array}$ & 0
\end{tabular}}\right),\cr
&& \Gamma^{i,11}=\left(\begin{array}{cc}
0 & -\Gamma^i\\ -\tilde\Gamma^i & 0\end{array}\right)=\left(
\mbox{\begin{tabular}{c|c}
0 & $\begin{array}{cc} 0 & -\gamma^i\\ -\bar\gamma^i & 0\end{array}$\\
\hline
$\begin{array}{cc} 0 & -\gamma^i\\ -\bar\gamma^i & 0\end{array}$ & 0
\end{tabular}}\right),\cr
&& \Gamma^{9,11}=\left(\begin{array}{cc}
0 & -\Gamma^9\\ -\tilde\Gamma^9 & 0\end{array}\right)=\left(
\mbox{\begin{tabular}{c|c}
0 & $\begin{array}{cc} -{\bf 1} & 0\\ 0 & {\bf 1}\end{array}$\\
\hline
$\begin{array}{cc} -{\bf 1} & 0\\ 0 & {\bf 1}\end{array}$ & 0
\end{tabular}}\right),
\end{eqnarray}
where $i=1,2,\dots,8$ and $\Gamma^{0i}$, $\Gamma^{09}$, $\Gamma^{i,11}$
$\Gamma^{9,11}$ are symmetric, whereas $\Gamma^{ij}$, $\Gamma^{i9}$,
$\Gamma^{0,11}$, are antisymmetric. Besides, these matrices are real and,
as a consequence of Eq. (A5), obey the commutation relations of the
Lorentz algebra.

Under the action of the Lorentz group a $D=11$ Dirac spinor is
transformed as
\begin{equation}
\delta\theta=-\frac 14 \omega_{\mu\nu}\Gamma^{\mu\nu}\theta.
\end{equation}
Since $\Gamma^{\mu\nu}$ matrices are real, the reality condition
$\theta^*=\theta$ is compatible with (A.10) which defines a
Majorana spinor. To construct Lorentz-covariant bilinear
combinations, one can note that
\begin{eqnarray}
\delta\bar\theta=+\frac 14\omega_{\mu\nu}
\bar\theta\Gamma^{\mu\nu},\qquad
\bar\theta\equiv\theta^{\rm T}C.
\end{eqnarray}
Then the combination $\bar\psi\Gamma^\mu\theta$ is a vector under the
action of the $D=11$ Lorentz group,
\begin{equation}
\delta(\bar\psi\Gamma^\mu\theta)={\omega^\mu}_\nu
(\bar\psi\Gamma^\nu\theta).
\end{equation}
The following properties are also useful
\begin{eqnarray}
&& \bar\psi\Gamma^{\mu_1}\cdots\Gamma^{\mu_k}\phi
={(-1)}^k\bar\phi\Gamma^{\mu_k}\cdots\Gamma^{\mu_1}\psi \cr               
&& \bar\psi\Gamma^{\mu_1\cdots\mu_k}\phi
={(-1)}^{\frac{k(k+1)}{2}}\bar\phi\Gamma^{\mu_1\cdots\mu_k}\psi.    
\end{eqnarray}

It is possible to decompose a $D=11$ Majorana spinor in terms of its
$SO(1,9)$ and $SO(8)$ components. Namely, it follows from Eq. (A.8) 
that in the decomposition
\begin{equation}
\theta=(\bar\theta_\alpha, \theta^\alpha), \qquad
\alpha=1,2\cdots 16
\end{equation}
$\theta$ and $\bar\theta$ are Majorana--Weyl spinors of opposite
chirality with respect to the $SO(1,9)$ subgroup of the $SO(2,9)$ group.
It follows from the third equation (A8) that in the decomposition
\begin{equation}
\theta=(\theta_a,\bar\theta'_{\dot a},\theta'_a,\bar\theta_{\dot a}),
 \qquad a,\dot a=1,\dots,8,
\end{equation}
the pairs $\theta_a$, $\theta'_a$ and
$\bar\theta'_{\dot a}$, $\bar\theta_{\dot a}$ are $SO(8)$ spinors of
opposite chirality.

It is convenient to define the $D=11$ light-cone $\Gamma$-matrices
\begin{eqnarray}
&& \Gamma^+=\frac 1{\sqrt 2}(\Gamma^0+\Gamma^9)=
\sqrt 2\left(\mbox{\begin{tabular}{c|c}
0 & $\begin{array}{cc} {\bf 1}_8 & 0\\ 0 & 0\end{array}$\\
\hline
$\begin{array}{cc} 0 & 0\\ 0 & {\bf 1}_8\end{array}$ & 0
\end{tabular}}\right),\cr
&& \Gamma^-=\displaystyle\frac 1{\sqrt 2}(\Gamma^0-\Gamma^9)=
\sqrt 2\left(\mbox{\begin{tabular}{c|c}
0 & $\begin{array}{cc} 0 & 0\\ 0 & {\bf 1}_8\end{array}$\\
\hline
$\begin{array}{cc} {\bf 1}_8 & 0\\ 0 & 0\end{array}$ & 0
\end{tabular}}\right),\cr
&& \Gamma^i=\left(\begin{array}{cc} 0 & \Gamma^i\\
-\tilde\Gamma^i & 0\end{array}\right), \cr
&& \Gamma^{11}=\left(\begin{array}{cc} {\bf 1}_{16} & 0\\
0 & -{\bf 1}_{16}\end{array}\right),
\end{eqnarray}
Then the equation $\Gamma^+\theta=0$ has a solution
\begin{equation}
\theta=(\theta_a,0,0,\bar\theta_{\dot a}).
\end{equation}
Besides, under the condition $\Gamma^+\theta=0$ the following
identities:
\begin{equation}
\bar\theta\Gamma^+\partial_1\theta=\bar\theta\Gamma^i\partial_1\theta=
\bar\theta\Gamma^{10}\partial_1\theta=0, \qquad
 (\bar\theta\Gamma^\mu\partial_1\theta)\Gamma^\mu\theta=0,
\end{equation}
hold.

\end{document}